\documentstyle[multicol,graphicx,aps,epsf]{revtex}

\begin{document}

\twocolumn[\hsize\textwidth\columnwidth\hsize\csname@twocolumnfalse%
\endcsname
\draft
\title{Nonlinear equation for anomalous diffusion: unified
power-law and stretched exponential exact solution}
\author{ L. C. Malacarne, R. S. Mendes, I. T. Pedron}

\address{Departamento de F\'\i sica, Universidade Estadual de Maring\'a,
Avenida Colombo 5790,\\ 87020-900,
 Maring\'a-PR, Brazil}
\author{  E. K. Lenzi}
\address{Centro Brasileiro de Pesquisas F\'\i sicas,
R. Dr. Xavier Sigaud 150, \\ 22290-180 Rio de Janeiro-RJ, Brazil }

\date{\today}

\maketitle

\begin{abstract}
The nonlinear diffusion equation $\frac{\partial \rho}{\partial
t}=D \tilde{\Delta} \rho^\nu$ is analyzed here, where
$\tilde{\Delta}\equiv \frac{1}{r^{d-1}}\frac{\partial}{\partial r}
r^{d-1-\theta} \frac{\partial}{\partial r}$, and $d$, $\theta$ and
$\nu$ are real parameters. This equation unifies the anomalous
diffusion equation on fractals ($\nu =1$) and the spherical
anomalous diffusion for porous media ($\theta=0$). Exact
point-source solution is obtained, enabling us to describe a large
class of subdiffusion ($\theta > (1-\nu)d$), normal diffusion
($\theta= (1-\nu)d$) and superdiffusion ($\theta < (1-\nu)d$).
Furthermore, a thermostatistical basis for this solution is given
from the maximum entropic principle applied to the Tsallis
entropy.
\end{abstract}
]

One of the most ubiquitous processes in nature is the diffusive
one. In this context, anomalous diffusion has awakened great
interest nowadays, in particular in a variety of physical
applications. A representative set of such applications of current
interest is surface growth and transport of fluid in porous
media\cite{spohn93}, diffusion in plasmas\cite{Berryman},
diffusion on fractals\cite{Stephenson}, subrecoil laser
cooling\cite{Bardou}, CTAB micelles dissolved in salted
water\cite{Ott}, two dimensional rotating flow\cite{Solomon} and
anomalous diffusion at liquid surfaces\cite{Bychuk}. The anomalous
diffusive process is commonly characterized from the mean-square
displacement time dependence, $\langle r^2 \rangle \propto
t^\sigma$, with $\sigma\neq 1$, i.e., we have superdiffusion for
$\sigma
>1$ and subdiffusion  for $\sigma <1$.
For a system which presents anomalous spreading, it is generally
associated to a non-gaussian space-time distribution, like
power-law or stretched exponential. In this framework, it is
desirable to incorporate, in a unified way, these two behaviours,
since it enables us to describe a wide class of diffusive
processes. The present work is dedicated to giving such unified
description.

Power-law or stretched exponential distributions  arise naturally
from  generalizations of the $d$-dimensional diffusion equation
\begin{equation}
\label{bd1}
 \frac{\partial \rho}{\partial t}=D \Delta \rho,
\end{equation}
with $\rho = \rho (\bar{x},t)$, $\bar{x}=(x_1, x_2, \ldots ,
x_d)$, $\Delta = \sum_{n=1}^d \partial^2/\partial x_n^2$, and $D$
being the diffusion coefficient. The nonlinear equation
\begin{equation}
\label{bd2}
 \frac{\partial \rho}{\partial t}=D \Delta \rho^\nu
\end{equation}
is just one of these generalizations, where $\nu$ is a real
parameter. Eq. (\ref{bd2})  has been employed to model diffusion
in porous medium (see Ref. \cite{spohn93} and references therein)
and in connection with generalized Tsallis
statistics\cite{plastino95,tsallis96,lisa}. Another important kind
of anomalous diffusion, in a tridimensional space, is related to
turbulent diffusion in the atmosphere and is usually
 described by\cite{richardson26}
\begin{equation}
\label{bd2a}
 \frac{\partial \rho}{\partial t}=\nabla \cdot ( K \nabla
 \rho ) \, ,
\end{equation}
where $K\propto r^{4/3}\, (r=|\bar{x}|)$. In a more general case,
we consider Eq. (\ref{bd2a}) in a $d$-dimensional space with
$K\propto r^{-\theta}$, where $\theta$ is a real parameter. Thus,
$\nabla \cdot ( K \nabla )$ is proportional to
$r^{-(d-1)}\frac{\partial}{\partial r} r^{d-1-\theta}
\frac{\partial}{\partial r} + A/r^{2-\theta}$ ($A$ is an operator
depending on the angular variables), and consequently
$\tilde{\Delta}\equiv r^{-(d-1)}\frac{\partial}{\partial r}
r^{d-1-\theta} \frac{\partial}{\partial r}$ is the radial part to
be considered in the study of the  spherical symmetrical solutions
of Eq. (\ref{bd2a}). In this context,  when $d$ is interpreted as
fractal dimension in an embedding $N$-dimensional space, the
equation
\begin{equation}
\label{bd3}
 \frac{\partial \rho}{\partial t}=D \tilde{\Delta} \rho
\end{equation}
has been used to study diffusion on fractals\cite{shaughnessy84}.

Here we are going  to propose the equation
\begin{equation}
\label{bd3a}
 \frac{\partial \rho}{\partial t}=\nabla \cdot ( K \nabla
 \rho^\nu ) \, ,
\end{equation}
as unification of Eqs. (\ref{bd2}) and (\ref{bd2a}). In fact, Eq.
(\ref{bd3a}) reduces to the correlated anomalous diffusion
(\ref{bd2}) if $K=D$, and to the generalized Richardson equation
(\ref{bd2a}) if $\nu=1$. The present study is mainly addressed to
the point-source solution of Eq. (\ref{bd3a}), because it
contains, as particular cases, a(n) (assimptotic) power-law and a
stretched exponential. In this away, we focus our attention on the
radial equation
\begin{equation}
\label{bd4}
 \frac{\partial \rho}{\partial t}= D\tilde{\Delta} \rho^\nu \, .
\end{equation}
Using this equation instead of Eq. (\ref{bd3a}) enables us to
analyze cases with noninteger $d$, so we can relate $d$ with a
fractal dimension. Therefore, in the following discussion, we are
going to consider $d$ as a nonnegative real parameter.

In order to motivate the ansatz  to obtain an exact time-dependent
solution for Eq. (\ref{bd4}), we recall the corresponding
solutions for equations Eqs. (\ref{bd1}), (\ref{bd2}), and
(\ref{bd3}). The time-dependent point-source solution for Eq.
(\ref{bd1}) is
\begin{equation}
\label{bd5} \rho(r,t)=\frac{\rho_0}{(4\pi
Dt)^{d/2}}\exp\left(-\frac{r^2}{4Dt}\right)\, ,
\end{equation}
where the normalization $\Omega_d \int_0^\infty \rho(r,t) r^{d-1}
dr=\rho_0$ and the $d$-dimensional solide angle $\Omega_d \equiv
2\pi^{d/2}/\Gamma(d/2)$   have been used. From Eq. (\ref{bd5}) we
 can easily  obtain the Einstein formula for the Brownian montion,
 {\it i.e.}, $\langle r^2 \rangle = 2dDt$.

 The analogous   solution for Eq. (\ref{bd2}) is\cite{plastino95,tsallis96,lisa}
\begin{equation}
\label{bd6} \rho (r,t) = \left[1- (1-q)\,\beta_1 (t)\,
r^2\right]^{\frac{1}{1-q}}\left/ Z_1(t)\right. ,
\end{equation}
where $q=2-\nu$, $Z_1 (t)\propto t^{d/[2+d(1-q)]}$  and $\beta_1
(t) \propto t^{-2/[2+d(1-q)]}$.
 It is
important to emphasize the short or long tailed shape of Eq.
(\ref{bd6}),  when compared with the normal diffusion (limit
$q\rightarrow 1$). When $q<1$ we have $\rho(r,t)=0$ for
$1-(1-q)\beta_1 (t) r^2 <0$, giving the short tailed behaviour for
$\rho(r,t)$. On the other hand, when $q>1$, the asymptotic
power-law behaviour for the solution (\ref{bd6}), $r^{-2/(q-1)}$,
shows that $\rho(r,t)$ is a long tailed function. This short or
long tailed behaviour for $\rho(r,t)$ reflects directly on the
mean-square displacement, leading to $\langle r^2 \rangle \propto
t^{2/[2+d(1-q)]}$. Again compared with the usual diffusion, $q=1$,
we have a superdiffusion (subdiffusion) for $q>1 (q<1)$.

The fundamental solution of Eq. (\ref{bd3}) is the stretched
exponential\cite{shaughnessy84}
\begin{equation}
\label{bd7} \rho (r,t) = \exp\left(-\beta_2 (t)\,
r^{\theta+2}\right)\left/ Z_2 (t)\right. ,
\end{equation}
which $Z_2(t) \propto t^{d/(\theta+2)}$ and $\beta_2 (t)\propto
t^{-1}$,  presenting a short (long) tailed behavior for $\theta >0
\;(\theta<0)$. Furthermore, the mean-square displacement behaviour
is $\langle r^2\rangle \propto t^{2/(\theta+2)}$. Thus, for
$\theta
>0$ ($\theta<0$) we have subdiffusive (superdifussive) regime.

Note that Eqs. (\ref{bd5}), (\ref{bd6}) and  (\ref{bd7}) can be
interpolated if we employ  a generalized stretched Gaussian
function, i. e., $G_{(q,\lambda)}(x)\equiv\left[1-(1-q)
|x|^\lambda\right]^{1/(1-q)}$ or $G_{(q,\lambda)}(x)\equiv 0$ when
$1-(1-q)|x|^\lambda <0$, with $q$ and $\lambda$ being real
parameters. In this direction, our ansatz to solve Eq. (\ref{bd4})
is
\begin{equation}
\label{bd9} \rho(r,t)=\left.
\left[1-(1-q)\,\beta(t)\,r^\lambda\right]^{\frac{1}{1-q}}\right/Z(t)
\end{equation}
or $\rho(r,t)=0$ if $1-(1-q)\beta(t)r^\lambda < 0$,where
$\beta(t)$ and $Z(t)$ are functions to be determined. By using
this ansatz in Eq. (\ref{bd4}) we verify that  $\beta(t)$ and
$Z(t)$, with $\lambda=\theta+2$ and $q=2-\nu$, obey the equations

\begin{eqnarray}
\label{bd10}
 \frac{dZ(t)}{dt} &=& D \lambda (2-q) d \beta(t)
Z^{q}(t) \nonumber
\\
\frac{d\beta(t)}{dt} &=& -D \lambda^2 (2-q) \beta^2(t) Z^{q-1}(t).
\end{eqnarray}

 The solutions of these nonlinear differential equations, that
 lead Eqs. (\ref{bd5}), (\ref{bd6}) and (\ref{bd7}) as limit cases
  of Eq. (\ref{bd9}), are
\begin{equation}\label{bd11}
\beta(t)= {\cal A}\; t^{-\frac{\lambda}{\lambda + d(1-q)}}
\,\,\,\,\,\,\,\,\;\;\;\;\;\;\; Z(t)={\cal B}\; t^{\frac{d}{\lambda
+ d(1-q)}}, \end{equation} where
\begin{eqnarray}\label{bd12}
 {\cal A}&=& \left\{\gamma^{q-1} \left[ \frac{}{} D\lambda (2-q)(\lambda
+d(1-q))\right]\right\}^{-\frac{\lambda}{\lambda + d(1-q)}}
\nonumber\\
 {\cal B}&=&\left\{\gamma \left[ \frac{}{} D\lambda (2-q)
(\lambda
+d(1-q))\right]^{\frac{d}{\lambda}}\right\}^{\frac{\lambda}{\lambda
+ d(1-q)}},
\end{eqnarray}
which
\begin{equation}\label{bd15}
\gamma = \frac{2\pi^{\frac{d}{2}}}{\lambda\rho_0}
\frac{\Gamma\left(\frac{d}{\lambda}
\right)}{\Gamma\left(\frac{d}{2} \right)} \left\{\begin{array}{cc}
 \frac{\Gamma\left(\frac{1}{q-1}-\frac{d}{\lambda}
\right)}{(q-1)^{\frac{d}{\lambda}}\Gamma\left(\frac{1}{q-1}
\right)} & (q>1)  \\\nonumber \\
  \frac{\Gamma\left(\frac{1}{1-q}+1
\right)}{(1-q)^{\frac{d}{\lambda}}\Gamma\left(\frac{d}{\lambda}+\frac{1}{1-q}+1
\right)} & (q<1)
\end{array}\right.
\end{equation}
 The normalization condition, $\Omega_d \int_0^\infty \rho(r,t) r^{d-1}dr=\rho_0$, employed
 in the above calculation can only be satisfied if  $\lambda >0$
and $ \lambda +d(1-q)>0$. From these conditions over the
parameters $q$ and $\lambda$, we verify that the exponents in
$\beta (t)$ and $Z(t)$ are respectively negative and positive. In
addition to the normalization condition, the restriction  $q<2$ is
necessary for $\rho (r,t)$ to be real. In the following, we assume
that the parameters obey the above restrictions.
 Of course,
by setting the appropriate limits of parameters
 $\theta$ and $\nu$, or equivalently $\lambda$ and $q$,
  the solutions (\ref{bd5}), (\ref{bd6}) and  (\ref{bd7})
are recovered, giving the full expression for $\beta_1 (t)$, $Z_1
(t)$, $\beta_2 (t)$ and $Z_2 (t)$.

By using the above solution we can calculate the  mean value of
$r^\alpha$; it is
\begin{equation}\label{bd16a}
 \langle r^\alpha\rangle = \frac{ \int_0^\infty r^\alpha \rho(r,t) r^{d-1}
dr}{ \int_0^\infty  \rho(r,t) r^{d-1}dr }={\cal C}_{\alpha} \;\;
 t^{\frac{\alpha}{\lambda+ d(1-q)}},
\end{equation}
where
\begin{eqnarray}\label{bd16b}
 {\cal C}_{\alpha}= {\cal A}^{-\frac{\alpha}{\lambda}}\frac{ \Gamma\left( \frac{d+\alpha}{\lambda}
 \right)}{\Gamma\left( \frac{d}{\lambda} \right)}
\left\{\begin{array}{cc}
    \frac{\Gamma\left(\frac{1}{q-1}- \frac{d+\alpha}{\lambda}
 \right)}{(q-1)^{\frac{\alpha}{\lambda}} \Gamma\left(\frac{1}{q-1}- \frac{d}{\lambda}
 \right)}& (q>1) \\\nonumber  \\
   \frac{\Gamma\left( \frac{d}{\lambda} +\frac{1}{1-q}+1\right) }
 { (1-q)^{\frac{\alpha}{\lambda}}
  \Gamma\left( \frac{d+\alpha}{\lambda}+\frac{1}{1-q}+1 \right)}& (q<1)
  \end{array}\right. ,
\end{eqnarray}
with ${\cal A}$ given  by Eq. (\ref{bd12}). When $q<1$, the mean
value $\langle r^\alpha \rangle$ always exist. On the other hand,
the existence of $\langle r^\alpha \rangle$  for $q>1$ impose a
further  restriction  over the parameters: $\lambda +d(1-q)>
\alpha (q-1)$.

\begin{figure}
 \centering
 \DeclareGraphicsRule{ps}{eps}{*}{}
 \includegraphics*[width=9cm, height=5cm,trim=5cm 2.5cm 1cm 4.5cm]{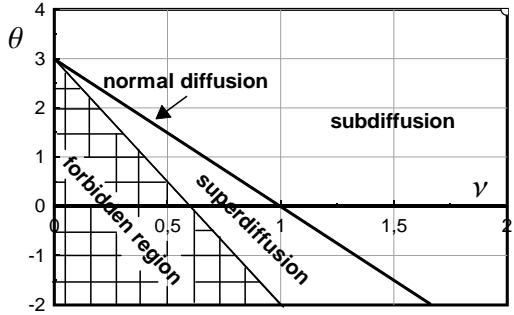}
  \caption{The diagram indicates the  diffusive regime
related to Eq. (\ref{bd4}) in terms of its parameters $\theta$ and
$\nu$ to $d=3$. Thus, by using $\langle r^2 \rangle \propto
t^{2/[\theta+2 +d(\nu-1)]}$ from Eq. (\ref{bd16a}) we have
classified the subdiffusive ($\theta
> (1-\nu)d$), normal ($\theta = (1-\nu)d$), and superdiffusive
($\theta < (1-\nu)d$) regimes. The forbidden region refers to the
region of parameters where  $\langle r^2 \rangle$ does not exist
(diverges). } \label{fig1}
\end{figure}

To decide if the diffusion is anomalous or normal, we consider
(\ref{bd16a})  with $\alpha=2$. In this way, we have $<r^2>\propto
t^\sigma$ with $\sigma=2/[2+\theta+ d(\nu-1)]$.
 Thus, the condition for normal diffusion, $\sigma=1$,
can be satisfied even when $\rho$ does not obey Eq. (\ref{bd1}),
i.e., $\theta=d(1-\nu)$ with $\theta\neq 0$ and $\nu \neq 1$. In
this case, we can also verify that the anomalous diffusive regime
induced by $\theta\neq 0$ is compensated by a convenient one with
$\nu\neq 1$. Furthermore, this competition between $\theta$ and
$\nu$ values can lead to a subdiffusion ($\sigma<1$) if $\theta
>d(1-\nu)$ or a supperdiffusion ($\sigma>1)$ if $\theta<
d(1-\nu)$. This classification is illustrated  in Fig.
(\ref{fig1}) for $d=3$.

\begin{figure}
 \centering
 \DeclareGraphicsRule{ps}{eps}{*}{}
 \includegraphics*[width=9cm, height=5.5cm,trim=1cm 0.5cm 1cm 1.5cm]{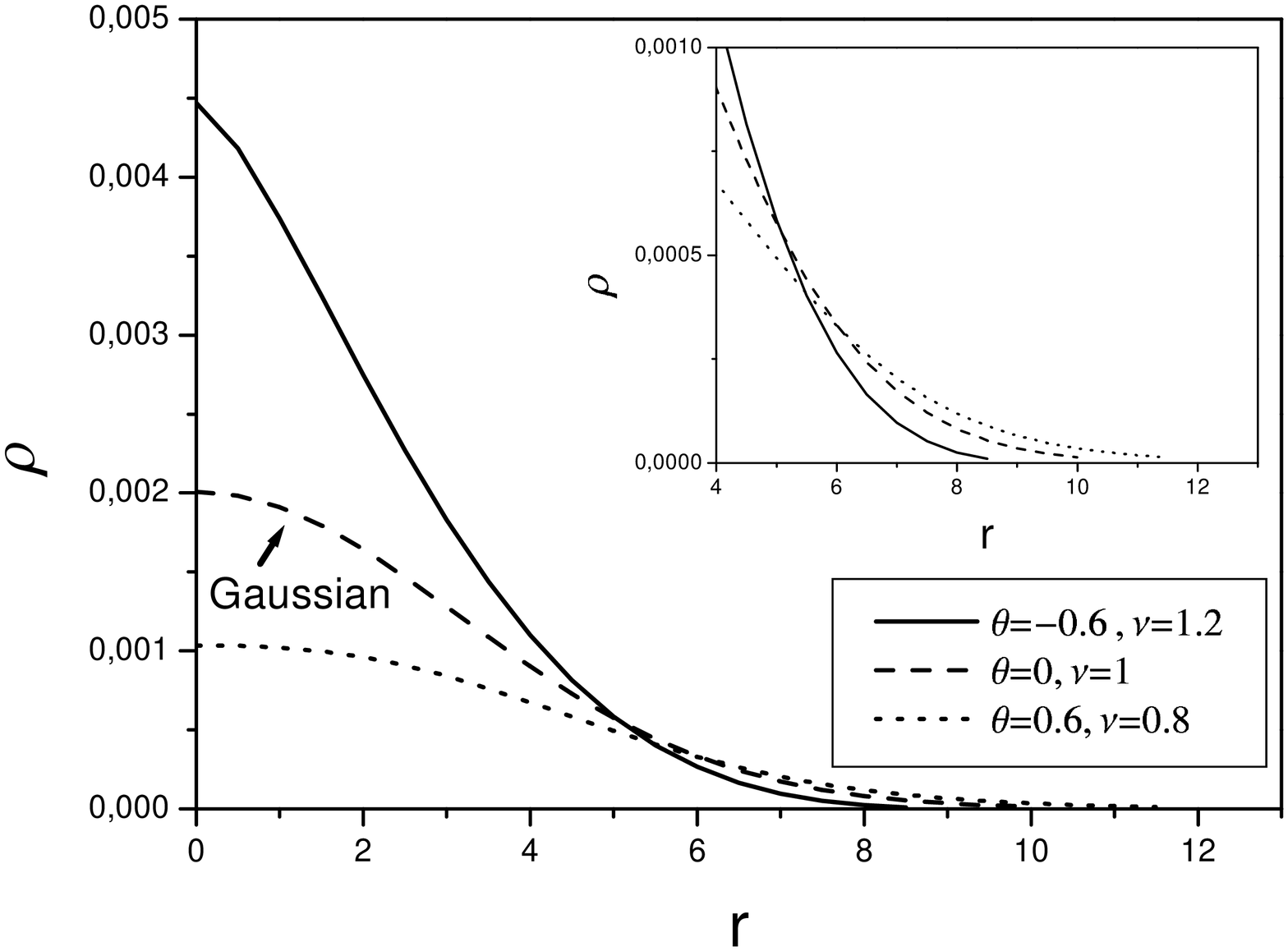}
 \caption{By considering the normal diffusion, $\theta = (1-\nu)d$,
 the shape of $\rho (r,t)$ with $d=3$ and $t=5$ is illustrated in
 three cases: short tail($\theta=-0.6$ and $\nu =1.2$),
 Gaussian ($\theta=0$ and $\nu=1$), and long tail ($\theta=0.6$ and
 $\nu =0.8$). Inset plot: detail of tail behaviour in the three cases above.
     } \label{fig2}
\end{figure}

In the following, we discuss the consequences of the above
classification on the $\rho$ shape. From the solution (\ref{bd6}),
for porous medium, and the solution (\ref{bd7}), for diffusion on
fractals, we can see that the superdiffusive (subdiffusive) regime
is associated to long (short) tail of $\rho (r,t)$ when compared
with Gaussian (\ref{bd5}). However, this conection is not valid in
general. To illustrate the relation between regime of diffusion
and tail behaviour of $\rho(r,t)$, we consider Fig. (\ref{fig2}).
In this figure, we plot $\rho(r,t)$ given by Eq. (\ref{bd9})
versus $r$ to some values of $\theta$ and $\nu$, subject to the
restriction $\theta=d(1-\nu)$ (the normal diffusion line indicated
in Fig. (\ref{fig1})). In this case we observed  short and long
tail behaviours compared with the Gaussian one.

To conclude our discussion about Eq. (\ref{bd3a}) and its radial
time-dependent solution (\ref{bd9}), we present an entropic basis
for this solution. This basis is motivated by  the Tsallis
generalized statistical mechanics\cite{T88,CT91,NORMA}, where the
Tsallis entropy\cite{T88} $S_q = (1- \sum_{j=1}^W p_j^q)/(1-q)$
plays a central role with $\{p_j \}$ being the probabilities for
the $W$ states of the system, and $q\in\mathcal{R}$ being the
Tsallis index (by taking the limit $q\rightarrow 1$ we recover the
usual entropy $S_1 = - \sum_{j=1}^W p_j \ln p_j$). To understand
the entropic basis, for simplicity, let us consider the
maximization of $S_q =(1-\int_{-\infty}^{\infty} \rho(x)^q\,
dx)/(q-1)$ subject to the constraints\cite{T88,CT91}
$\int_{-\infty}^{\infty} \rho(x)\, dx=1$ and
$\int_{-\infty}^{\infty}|x|^\lambda \rho(x) \, dx=U_q$. This
maximization leads to $\rho(x)= [1-(1-q) \beta |x|^\lambda
]^{1/(1-q)}/{\cal Z}_q$, where  ${\cal
Z}_q=\int_{-\infty}^{\infty} [1-(1-q) \beta |x|^\lambda
]^{1/(1-q)} dx  $ and $\beta$ is related to the Lagrange
multipliers.

We thank CNPq and PRONEX (Brazilian agencies) for partial
financial support.

\end{document}